# On the Spatial Ambiguity Function

Arthur N. Yuryev

30th Central Scientific Research Institute, Ministry of Defence (USSR)

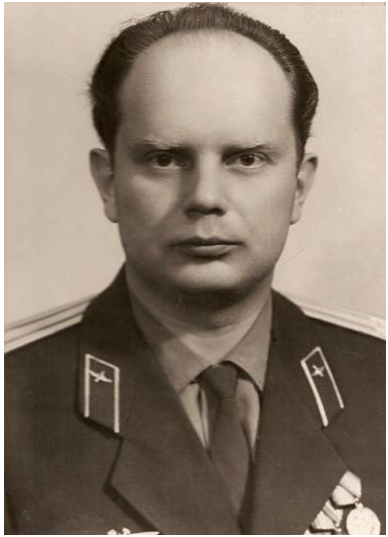

## Abstract

The ambiguity function of a spatial signal in the "signal spatial frequency displacement vs antenna linear displacement" coordinates is considered in this paper; in case of a monochromatic signal with known carrier frequency, spatial signal vector angular displacement is considered as the second coordinate. The analysis is made taking into account an external (reradiated or thermal) background which contributes considerable peculiarities in the properties of the spatial ambiguity function. Accuracy of measuring the signal linear or angular displacement and resolution are determined. Examples for the Fresnel and Fraunhofer zones are given.[1]

---

# Arthur Nikolaevich YURYEV

A. N. Yuryev was born at Rybinsk city, Russia, on July 5, 1932. He graduated in 1956 with honours and Engineer degree in Radar engineering, and received the Cand. Sci. degree in 1962 from Prof. N. E. Zhukovsky Air Force Engineering Academy, Moscow, USSR. Later, in 1976, he received the Dr. Sci. degree.

Senior Researcher and Professor ranks in Radar and Radionavigation were conferred to him in 1965 and 1984, respectively.

From 1950 he was with the USSR Air Force and from 1962 he worked at the 30th Central Scientific Research Institute of the Ministry of Defence of USSR also known as the Central Research Institute of Aviation and Space Technology.

The field of his scientific interests included, e.g., radar and navigation, image recognition, information theory, communication and control theory and their applications. He was an author (with coauthors) of a number of monographs such as "Correlation-Extremum Methods of Navigation" (Moscow, 1982) and "Adaptive Control Systems of Aircrafts" (Moscow, 1987), "Numerical methods of binary image processing and recognition" (Krasnoyarsk, 1992), etc.

Col. A. N. Yuryev was a member of the A. S. Popov All-Union Radio, Electronics and Communication Engineering Society.

Prof. A. N. Yuryev deceased on September 30, 1990.

# ON THE SPATIAL AMBIGUITY FUNCTION

A. N. Yuryev

*Abstract.* The ambiguity function of a spatial signal in the "signal spatial frequency displacement vs antenna linear displacement" coordinates is considered in this paper; in case of a monochromatic signal with known carrier frequency, spatial signal vector angular displacement is considered as the second coordinate. The analysis is made taking into account an external (reradiated or thermal) background which contributes considerable peculiarities in the properties of the spatial ambiguity function. Accuracy of measuring the signal linear or angular displacement and resolution are determined. Examples for the Fresnel and Fraunhofer zones are given[1].

---

**Introduction.**

The radar signal ambiguity function, introduced in the works [1,2], was the basis and starting point for a whole sequence of investigations of resolution and accuracy of radar measurements. The main peculiarity of this function is the property of invariance of uncertainty body cross section relative to signal shape, what is conditioned by that the signal parameters resolved - in the classical treatment they are time delay and carrier frequency - are connected with one another by a pair of Fourier transforms. Then the ideas of works [1,2] were propagated to the other parameters of signal and target (signal arrival direction, target angular velocity and acceleration and so on), which are not connected by Fourier transforms (e.g. [3-6]). Unified signal ambiguity functions characterizing processing system, which is an optimum on the background of correlated interferences, was also studied.

Despite a large number of works on fundamental and applied questions, in which, in one form or another, the concept of ambiguity function have been used (e.g. [7-11]), its ideas propagation to the region of space measurements have been limited by consideration of "unclassical" cases - by the questions of signal arrival direction measuring, target distance in near field region [10, 12-14] and so on. Attention has not been paid on the properties of the space ambiguity function (SAF) on the "processing system (antenna) linear displacement - signal space frequency displacement" plane (the displacement of a space frequency converts to angular displacement of processing system relative to radiating object in case of signal with known carrier frequency). This seemed to be connected with the fact that such measurements did not find practical application until recently. Appearing of correlation-external navigation systems [15,16] has attracted attention to principle possibility of mutual measuring of object linear and angular displacements, especially in cases when the image of radiating or reflecting background is formed simultaneously in direct form (as a result of using of multiple-beam or scanning diagram-forming structure) and in form of inverse background image on the processing system aperture. In this case SAF is that instrument, with the help of which the estimate of principle potentialities of such systems may be given. So the investigation of SAF properties on the "antenna linear displacement - signal space frequency displacement" plane appears to be rather actual problem, which is considered in this paper.

**1. Peculiarities of SAF for system working against radiating background.**

Let us consider the following problem, which is characteristic for radar measurements. Assume that the space signal $\dot{H}(\bar{r})$ comes from a radiation source extensive by angular coordinate and frequency into an antenna system, element coordinates of which are defined by a vector $\bar{r} \in \bar{R}_o$:

$$\dot{H}(\bar{r}) = \int_{\bar{\Psi}_o} \dot{\xi}_t(\bar{\psi}) exp(-2\pi i \bar{\psi}\bar{r}) d\bar{\psi} , \qquad (1)$$

where $\bar{\psi} = \bar{\rho}/\lambda$ is radiation space frequency vector ($\lambda$ is a wavelength, $\bar{\rho}$ is a unit vector directed from any point of antenna system to arbitrary radiating point of surface); integrating over vector $\bar{\psi}$ should be understood as integrating in two-dimensional region over orthogonal components of the vector ($d\bar{\psi}=d\psi_x d\psi_y$). Random function $\dot{\xi}_t(\bar{\psi})$, which is complex amplitude of radiated (reflected) signal, characterizes local peculiarities of given background part and contains useful information, which might be used for determination of relative space position of receiving aperture; subscript 't' designates availability of the complex amplitude time fluctuations.

The space signal $\dot{H}(\bar{r})$, laid on self distribution of receiving aperture field $\dot{h}(\bar{r})$, forms resulting signal, which inputs the signal space processing system and is a resulting field in antenna aperture:

$$\dot{H}(\bar{r})=\dot{h}(\bar{r})\dot{H}(\bar{r}). \qquad (2)$$

Optimal linear filter, fulfilling linear processing of the signal (2), has a complex weight characteristics for uncorrelated space interferences:

$$\dot{G}(\bar{r})=\dot{H}^*(\bar{r})exp\{-2\pi i\bar{\psi}_o\bar{r}\}, \qquad (3)$$

where $\bar{\psi}_o$ is expected vector of coming signal space frequency being an element of some multitude $\{\bar{\Psi}_{o1}\}$, which reflects the fact that the space processing system forms a continuum of received beams.

Normalized with respect to its maximum output effect of optimal space processing system comprises a stochastic SAF of the signal - stochastic character of the function is conditioned by chance cha-racter of the complex amplitude $\dot{\xi}_t(\bar{\psi})$ in time. Simmetrized form of ambiguity function with respect to parameters, which characterize space (linear) $\bar{\Delta}$ and space-frequency (angular) $\bar{\Psi}=\bar{\psi}-\bar{\psi}_o$ displace-ments of aperture with respect to some coordinate system, will be used below.

The stochastic SAF is:

$$\dot{\theta}_t(\bar{\Delta},\bar{\Psi}) = \int_{\bar{R}_0} \dot{H}(\bar{r}+\frac{\bar{\Delta}}{2})\dot{H}^*(\bar{r}-\frac{\bar{\Delta}}{2})exp\{2\pi i\bar{\Psi}\Phi\bar{r}\}d\bar{r} \ . \qquad (4)$$

Regarding that the field distribution in the antenna aperture

$$\dot{H}(\bar{r}) = \int_{\bar{\Psi}_0} \dot{F}(\bar{\psi})exp\{-2\pi i\bar{\psi}\bar{r}\}d\bar{\psi} \ , \qquad (5)$$

the stochastic SAF (4) may be presented as

$$\dot{\theta}_t(\bar{\Delta},\bar{\Psi}) = \int_{\bar{\Psi}_0} \dot{F}(\bar{\psi}+\frac{\bar{\Psi}}{2})\dot{F}^*(\bar{\psi}-\frac{\bar{\Psi}}{2})exp\{-2\pi i\bar{\psi}\Phi\bar{\Delta}\}d\bar{\psi} \ . \qquad (6)$$

Expressions (4) and (6) differs by immaterial phase factor, which is omitted here. Remark that the resulting directivity diagram of the space processing system $\dot{F}(\bar{\psi})$ may be considered as smoothed "image" of complex amplitude $\dot{\xi}_t(\bar{\psi})$ random distribution. Really, substituting the expression (2) into the expression

$$\dot{F}(\bar{\psi}) = \int_{\bar{R}_0} \dot{H}(\bar{r})exp\{2\pi i\bar{\psi}\bar{r}\}d\bar{r} \qquad (7)$$

regarding (1) we obtain:

$$\dot{F}(\bar{\psi}_1) = \int_{\bar{\Psi}_0} \dot{\xi}_t(\bar{\psi})\dot{f}(\bar{\psi}_1-\bar{\psi})d\bar{\psi} \ , \qquad (8)$$

where the function

$$\dot{f}(\bar{\psi}) = \int_{\bar{R}_0} \dot{h}(\bar{r})exp\{2\pi i\bar{\psi}\bar{r}\}d\bar{r} \qquad (9)$$

represents a directivity diagram of the system, which is formed by own distribution of antenna field $\dot{h}(\bar{r})$.

We shall assume that time fluctuations of different points of radiating surface are uncorrelated, i.e.

$$<\dot{\xi}_t(\bar{\psi})\dot{\xi}_t^*(\bar{\psi}_1)>_t = \dot{\sigma}(\bar{\psi})\dot{\sigma}^*(\bar{\psi}_1)\delta(\bar{\psi}-\bar{\psi}_1). \tag{10}$$

The angular brackets with subscript 't' designate statistical averaging over time fluctuations. The function $\dot{\sigma}(\bar{\psi})$ has quasi-random character and defines radiation intensity in given space-frequency zone. Averaging the stochastic SAF (4) regarding (1), (2) and (3), we have the following expression:

$$\theta(\bar{\Delta},\bar{\Psi}) = |\dot{\theta}(\bar{\Delta},\bar{\Psi})| = |<\dot{\theta}_t(\bar{\Delta},\bar{\Psi})>_t| = \theta_O(\bar{\Delta},\bar{\Psi}) K_b(\bar{\Delta}), \tag{11}$$

where

$$\theta_O(\bar{\Delta},\bar{\Psi}) = \left| \int_{\bar{R}_O} \dot{h}(\bar{r}+\frac{\bar{\Delta}}{2})\dot{h}^*(\bar{r}-\frac{\bar{\Delta}}{2}) exp\{2\pi i\bar{\Psi}\Phi\bar{r}\} d\bar{r} \right| =$$

$$= \left| \int_{\bar{\Psi}_O} \dot{f}(\bar{\psi}+\frac{\bar{\Psi}}{2})\dot{f}^*(\bar{\psi}-\frac{\bar{\Psi}}{2}) exp\{-2\pi i\bar{\psi}\Phi\bar{\Delta}\} d\bar{\psi} \right|; \tag{12}$$

$$K_b(\bar{\Delta}) = \left| \int_{\bar{\Psi}_O} |\dot{\sigma}(\bar{\psi})| exp\{-2\pi i\bar{\psi}\Phi\bar{\Delta}\} d\bar{\psi} \right|. \tag{13}$$

We shall assume below without changing of designations that average SAF (11) is normalized so that $\theta(0,0)=1$.

The function $\theta_O(\bar{\Delta},\bar{\Psi})$ represents SAF of antenna aperture proper. This function characterizes uncertainty between antenna aperture linear displacement on the background of unit point radiator, which is placed in the region $\bar{\Psi}_O$, in coordinates connected with this radiator and antenna angular position with respect to this radiator. The function $K_b(\bar{\Delta})$ is a modulus of correlation function of field distribution in antenna aperture, which is formed by external background. Really, the function $|\dot{\sigma}(\bar{\psi})|$ may be considered as energetic spectrum of radiating (reflecting) background formed by some aperture $\bar{R}_1$. In this case

$$\dot{\sigma}(\bar{\psi}) = \int_{\bar{R}_1} \dot{h}_b(\bar{r}_1) exp\{2\pi i \bar{\psi}\bar{r}_1\} d\bar{r}_1 \ , \qquad (14)$$

and the field distribution, which is formed by radiating background (background radio-holography information component), in the aperture after averaging of (11) is

$$\dot{h}_b(\bar{r}) = \int_{\bar{\Psi}_0} \dot{\sigma}(\bar{\psi}) exp\{2\pi i \bar{\psi}\bar{r}\} d\bar{\psi} \ , \qquad (15)$$

and the function

$$K_b(\bar{\Delta}) = \left| \int_{\bar{R}_0} \dot{h}_b(\bar{r}+\frac{\bar{\Delta}}{2}) \dot{h}_b^*(\bar{r}-\frac{\bar{\Delta}}{2}) d\bar{r} \right| . \qquad (16)$$

So the SAF structure, when working against the background of radiating (or reflecting) surface - and namely in this case SAF has a practical significance,- differs considerably from the structure of classical frequency-time ambiguity function [1,2]. In case of SAF the correlation function $K_b(\bar{\Delta})$ of averaged distribution of the field formed by intensity relief of radiating background, on which the system works, is to be allowed for. In this connection the other characteristics of the considered SAF also differ from those of classical ambiguity functions [1,2]. The volume of SAF uncertainty body equals

$$V = \int\int_{\bar{R}_0 \bar{\Psi}_0} \theta^2(\bar{\Delta},\bar{\Psi}) d\bar{\Delta} d\bar{\Psi} = \int_{\bar{R}_0} K_0(\bar{\Delta}) K_b^2(\bar{\Delta}) d\bar{\Delta} \qquad (17)$$

where

$$K_0(\bar{\Delta}) = \int_{\bar{R}_0} \left|\dot{h}(\bar{r}+\frac{\bar{\Delta}}{2})\right|^2 \left|\dot{h}(\bar{r}-\frac{\bar{\Delta}}{2})\right|^2 d\bar{r}, \qquad (18)$$

and in doing so in connection with normalization accepted

$$\int_{\bar{R}_0} \left|\dot{h}(\bar{r})\right|^2 d\bar{r} = 1, \qquad (19)$$

and hence

$$\int_{\bar{R}_0} K_0(\bar{r}) d\bar{\Delta} = 1. \qquad (20)$$

As $K_b(0)=1$, so in this case, unlike the classical frequency-time ambiguity function,

$$V \leq 1, \qquad (21)$$

and $V=1$ when $K_b(\bar{\Delta})=1$ what takes place when the system works on the background of point radiator. This is natural, that the volume of uncertainty body of the field distribution itself in antenna aperture equals unity.

It follows from (11) that the functions $K_b(\bar{\Delta})$ and $\theta_0(\bar{\Delta},\bar{\Psi})$ make a contribution in resolution over linear opening of the aperture, which "illuminates" the background, is considerably less in the

- 8 -

di-rection of vector $\bar{\Delta}$ of the space processing system aperture R, the system resolution over vector $\bar{\Delta}$ is determined by the function $K_b(\bar{\Delta})$, and the width of correlation function $K_b(\bar{\Delta})$ having an order of the radiating aperture size. A radar system with synthetic aperture may be an example illustrating this situation. The resolution of such system over linear coordinate is determined by physical size of a real radiating aperture, while receiving (synthetic) aperture - which determines the resolution over space frequency - considerably excel radiating one in its sizes.

Let us illustrate the essence of the results obtained.

Uncertainty in measuring of linear displacement of some aperture - e.g. situated on moving object - and in meaning of observed (ra-diating) object space frequency is explained in Fig.1. When using the signal with known carrier, the mentioned uncertainty has a sense of uncertainty between antenna linear displacement and space signal angular displacement. Three space locations of antennas (space sig-nal) $A_1$, $A_2$ and $A_3$ depicted, every of which differ from the others at least by one of the coordinates - linear x or angular $\psi_x$, may be resolved from each other only at certain properties of the space si-gnal. The case is shown in Fig.1a, when the frequency-space radiation spectrum $|\dot{\sigma}(\psi_x)|$ is not wide enough; the system resolution over x and $\psi_x$ coordinates is determined by the SAF cross-section given in the same picture. The other situation is depicted in Fig.1b, where the background frequency-space spectrum has wider extent, so higher resolution over x-coordinate ($\delta_\Delta^{''} < \delta_\Delta^{'}$) is provided, the system space frequency resolution remaining unchanged. Thus, allowing for the external space-frequency distribution of radiating background considerably changes the properties of ambiguity function for the space signal in comparison with the classical one in "time delay vs carrier frequency shift" coordinates, where resolution improvement over one parameter results in its deterioration over another one.

## 2. SAF for radiation near field region.

Dwell now on SAF of signal detecting from near field region. The induced field distribution in apparatus (1) may be presented in this case as

$$H(\bar{r}) = \int_{\bar{\Psi}_o}^{\Phi} \xi_t(\bar{\psi}) exp\{-2\pi i(\bar{\psi}\bar{r} - \frac{a}{2}|\bar{r}|^2)\}d\bar{\psi} \qquad (22)$$

where the parameter $a = \frac{\bar{n}\bar{\Phi}\bar{\psi}}{D} = \frac{\bar{n}\bar{\rho}}{\lambda D}$ ($n$ is a unit vector normal to the ap-paratus surface, D means the distance from apparatus to radiating surface). Let us assume that $\bar{n}\bar{\rho} \cong 1$, i.e. the radiating area is located in a narrow angular zone adjoining vector $\bar{n}$, that as a rule takes place in practice.

Acting by analogy with Section 1, we obtain the following expression for SAF, which is analogous to formula (11):

$$\theta(\bar{\Delta},\bar{\Psi}) = \theta_o^n(\bar{\Delta},\bar{\Psi})K_b(\bar{\Delta}); \qquad (23)$$

in (23), the ambiguity function for point radiation situated in near field region is

$$\theta_o^n(\overline{\Delta},\overline{\Psi}) = \left| \int\limits_{\overline{R}_o} \dot{h}(\overline{r}+\frac{\overline{\Delta}}{2})\dot{h}^*(\overline{r}-\frac{\overline{\Delta}}{2})exp\{2\pi i(\overline{\Psi}+a\overline{\Delta})\overline{r}\}d\overline{r} \right| =$$

(24)

$$= \left| \int\limits_{\overline{\Psi}_o} \dot{f}(\overline{\psi}+\frac{\overline{\Psi}+a\overline{\Delta}}{2})\dot{f}^*(\overline{\psi}-\frac{\overline{\Psi}+a\overline{\Delta}}{2})exp\{-2\pi i\overline{\psi}\Phi\overline{\Delta}\}d\overline{\psi} \right|.$$

The important relation results from comparison (12) and (24), which connects SAF for far and near fields of radiation for point signal source:

$$\theta_o^n(\overline{\Delta},\overline{\Psi}) = \theta_o(\overline{\Delta},\overline{\Psi}+a\overline{\Delta}).$$  (25)

In particular, at uniform proper field distribution in the opening of angular and linear aperture of the length L we have for far field:

$$\theta_o(\overline{\Delta}_x,\overline{\Psi}_x) = \left| \frac{sin\{(1-|\Delta_x|/L)\pi\Psi_x L\}}{\pi\Psi_x L} \right|;$$  (26)

we have under the same conditions for near field:

$$\theta_o^n(\overline{\Delta}_x,\overline{\Psi}_x) = \theta_o(\overline{\Delta}_x,\overline{\Psi}_x+a\overline{\Delta}_x) = \left| \frac{sin\{(1-|\Delta_x|/L)\pi(\Psi_x L+\Delta_\Psi\Delta_x)\}}{\pi(\Psi_x L+\Delta_\Psi\Delta_x)} \right|$$  (27)

where $\Delta_x$ and $\Psi_x$ are respectively a linear shift along x-axis, on which apparatus is located, and a shift of the space frequency vector projection on this axis; $\Delta_\Psi = aL$ is the space frequency deviation along apparatus. Relations (26) and (27) are analogous to corresponding time ambiguity functions for rectangular pulses with unmodulated and frequency-modulated carrier. The function $K_b(\overline{\Delta})$ in (23) is determined by the same relations as in case when the far field radiation has been considered ((13) and (16)).

The SAF cross-sections are presented in Figs.2 and 3 for far (Fig.2) and near (Fig.3) radiation fields at linear antenna aperture. The cross-sections of (26) and (27) functions are also plotted there. When the function $K_b(\bar{\Delta})$ was calculated, the energy distribution $|\dot{\sigma}(\psi_x)|^2$ was used, which is depicted in Fig.4a; the $K_b(\Delta_x)$ function is plotted in Fig.4b.

### 3. Resolution and accuracy of measuring of spatial parameters of a signal.

The expressions (11) and (23) allow calculating the potential resolution over parameters $\bar{\Delta}$ and $\bar{\Psi}$, determining linear and space-frequency displacement of one object with respect to another - extensive radiating object - and also interdependence of these parameters measurements. In particular, the resolution over $\Delta_x$ and $\Psi_x$ parameters may be derived for the system with linear aperture from the following relations:

$$\delta_\Delta^{-2} = -\left.\frac{\partial^2 \theta(\Delta_x, \Psi_x)}{\partial \Delta_x^2}\right|_{\substack{\Delta_x=0 \\ \Psi_x=0}} \; ; \; \delta_\Psi^{-2} = -\left.\frac{\partial^2 \theta(\Delta_x, \Psi_x)}{\partial \Psi_x^2}\right|_{\substack{\Delta_x=0 \\ \Psi_x=0}} . \qquad (28)$$

The potential precision of non-mutual measurements is determined by the system resolution. The dispersions of non-mutual estimates $\sigma_{o\Delta}^2$ and $\sigma_{o\Psi}^2$ of the parameters $\Delta_x$ and $\Psi_x$ are expressed via relations (28):

$$\sigma_{o\Delta}^{-2} = q\delta_\Delta^{-2} \; ; \; \sigma_{o\Psi}^{-2} = q\delta_\Psi^{-2}. \qquad (29)$$

where q is energetic signal-to-noise ratio [17].

The dispersions of mutual estimates $\sigma_\Delta^2$ and $\sigma_\Psi^2$ are connected with the dispersions of non-mutual estimates in the following way [17]:

$$\frac{\sigma^2_{o\Delta}}{\sigma^2_{\Delta}} = \frac{\sigma^2_{o\Psi}}{\sigma^2_{\Psi}} = 1-r^2_{\Delta\Psi} \qquad (30)$$

where correlation coefficient of mutual measurements $r_{\Delta\Psi}$ may be calculated from the formula

$$r_{\Delta\Psi} = \left.\frac{\partial^2 \theta(\Delta_x,\Psi_x)}{\partial\Delta_x\partial\Psi_x}\right|_{\substack{\Delta_x=0\\\Psi_x=0}} \delta_{\Delta}\delta_{\Psi} . \qquad (31)$$

The relations mentioned above determine the potentialities of resolution and measuring of $\Delta_x$ and $\Psi_x$ parameters for the case when mutual shift of all elements of the weight function (3) and the space signal (2) takes place. This case corresponds to the variant of "binding" of proper distribution of reference signal aperture (weight function) to the radiating object, and the destinations in the space location of proper distribution in (2) and (3) contribute to the total resolution of the system over objects linear shift with respect to each other.

The case, when the spatial signal $\dot{H}(\bar{r})$ induced by radiating object has its own additional shift with respect to antenna aperture, is of interest. SAF for this case (23) may be presented by the expression

$$\theta(\bar{\Delta},\bar{\Delta}_1,\bar{\Psi})=\theta^n_o(\bar{\Delta},\bar{\Psi}+a\bar{\Delta}_o)K_b(\bar{\Delta}_o), \qquad (32)$$

where $\bar{\Delta}_o=\bar{\Delta}+\bar{\Delta}_1$. When a=0 we have the corresponding expression for radiation far field (11).

The alterative, when the shift of apertures (proper field distribution) $\bar{\Delta}$ is absent in the expression (32), is of the greatest practical significance. It corresponds to the case, when (e.g. in correlation-extremum systems) the current and standard apertures are superimposed exactly on one another and the shift of received field occurs, which is conditioned either by carrier movement or by standards change. In doing so, it is necessary to substitute the function (32) at $\bar{\Delta}=0$ for $\theta(\Delta_x, \Psi_x)$ in the expressions (28) and (31) and to make differentiation over $\Delta_{1x}$ parameters instead of $\Delta_x$. Remark that the relations obtained in such a way will also describe the case when the weight function (3) is formed on the aperture of less size than the space signal (2), and when the maximum of correlation function $K_b(\bar{\Delta}_1)$ modulus is being found, it as if "slides" along the larger aperture.

The results of analysis of antenna systems of both mentioned types - Variant 1 corresponds to using of the function (23), Variant 2 corresponds to the function (32) - are brought together in Table 1.

The following designations are used in Table 1.

$$l_x = \left[ \int_{\bar{R}_O} x^2 |\dot{h}(x)|^2 dx - \left( \int_{\bar{R}_O} x |\dot{h}(x)|^2 dx \right)^2 \right]^{1/2}$$

is a root-mean square width of proper field distribution on the aperture;

$$l_{\psi f} = \left[ \int_{\bar{\Psi}_O} \psi_x^2 |\dot{f}(\psi_x)|^2 d\psi_x - \left( \int_{\bar{\Psi}_O} \psi_x |\dot{f}(\psi_x)|^2 d\psi_x \right)^2 \right]^{1/2}$$

is a root-mean square width of space-frequency spectrum of antenna;

$$l_{\psi\sigma} = \left[ \int_{\bar{\Psi}_O} \psi_x^2 |\dot{\sigma}(\psi_x)|^2 d\psi_x - \left( \int_{\bar{\Psi}_O} \psi_x |\dot{\sigma}(\psi_x)|^2 d\psi_x \right)^2 \right]^{1/2}$$

is a root-mean square width of space-frequency spectrum of extensive object radiation.

In the disignations given, the following normalization is used:

$$\int_{\overline{R}_O} |\dot{h}(x)|^2 dx = \int_{\overline{\Psi}_O} |\dot{f}(\psi_x)|^2 d\psi_x = \int_{\overline{\Psi}_O} |\dot{\sigma}(\psi_x)|^2 d\psi_x = 1.$$

Let us analyze the relations given in Table 1.

Note first of all that the distinction of Variant 1 relations from those of Variant 2 is in the following: the value $l^2_{\psi f}$ is additionally involved in all the expressions from Variant 1, into which the values determining the space-frequency spectrum width enter, this value widen the space spectrum total extent. However, the practical significance of Variant 1 is very limited, because the proper antenna field distribution can determine the relative shift between the objects only in exceptional cases (e.g. if the "reference" aperture is situated on a satellite constantly hanging over the radiating object). The Variant 2 is of practical importance for mobile objects; this is, in particular, one of basic distinctions of time and space measurements: it is known that for time measurements the corresponding analog of $l_{\psi f}$ quantity (signal spectrum width) is a factor determining the signal resolution over time delay. For the space measurements, the quantities $l_{\psi\sigma}$ and $al_x$, which are the space-frequency spectrum width being determined by radiating object and by radiation zone, plays the main role for object linear displacement estimating. Therewith the space signal linear resolution (a=0) does not depend on spectrum width $l_{\psi\sigma}$, i.e. it is independent of correlation function (16) of the fields $\dot{h}(x)$ induced on the aperture by radiating object; the narrower correlation function (16), i.e. the larger $l_{\psi\sigma}$ value, the more accurate the object position is detected. In case of mutual measurements of linear and space-frequency parameters this component of space-frequency spectrum width entirely determines the resolution over linear coordinate. In non-mutual measurements (when $\Psi_x$ quantity is known), the precision of linear displacement measurements is also influenced by the quantity $al_x$ determined by the spatial frequency deviation, what takes place in radiation near field - for rectangular field distribution in antenna aperture the spatial frequency deviation $\Delta_x = al_x/2\sqrt{3}$. The system resolution over linear displacement is entirely determined by the spectrum width $al_x$ for point radiation source ($l_{\psi\sigma}=0$).

The accuracy of estimation of space-frequency shift (i.e. of extended object image displacement over angular and frequency parameters) in non-mutual measurements is determined only by root-mean square width of proper field distribution in antenna aperture $l_x$. In mutual measurements, this dependency is more complicated. In particular, the exactness of this shift measuring is inverse proportional to entire width of radiation space-frequency spectrum $(l_{\psi\sigma}^2 + a^2 l_x^2)^{1/2}$ in radiation near field. It should not be forgotten that radiation space-frequency spectrum width is determined not only by receiving angular sector width, but also by radiating signal frequency bandwidth.

Dwell now on correlation coefficient for mutual measurements. The measurements of linear and space-frequency parameters are independent for radiation far field ($a=0$). However, this dependency increases with the growth of $al_x$ quantity - i.e. with decrease of ob-ject distance or radiation wavelength or with growth of effective aperture, the modulus of correlation coefficient tends in the limit to unity. The width of space-frequency radiating object spectrum affects the quantity of correlation coefficient : the narrower spectrum distribution $|\dot\sigma(\psi_x)|^2$, the higher correlation; the measurements of linear and space-frequency parameters become hard dependent for point radiation in near field region. Note also that correlation coefficient of measurements is negative - i.e. with increase of measurements error meanings of one of parameters, this of another parameter tends to change in negative direction.

In conclusion, we examine a brief example of error estimation using the above approach.

Let us consider radar with synthetic aperture: $\lambda$=10cm, antenna aperture width d=10m. A carrier space location is measured from the signal reflected by some lot of landscape situated at the distance D=10km; q=2. Let us estimate the limit potentialities of mutual measuring of the carrier linear location on the trajectory and its angular orientation (Table 1, Variant 2).

The maximum possible synthetic aperture length L=D$\Delta\varphi$ =100m. Assuming the system aperture function to be uniform we have $l_x$=L/2$\sqrt{3}$=29m. Suppose radiation space-frequency spectrum width to be some less than antenna space-frequency characteristics width $l_{\psi f}$=$\Delta\varphi/\lambda$=1/d ($l_{\psi\sigma}$=(2$\sqrt{3}$d)$^{-1}$=2.9×10$^{-2}$ 1/m). We obtain under this conditions: $r_{\Delta\psi}$=0.7; $\sigma_\Delta$=3.8m; $\sigma_\psi$=5.4×10$^{-3}$ 1/m; the last value corresponds to root-mean square of angular error $\sigma_\varphi$=$\lambda\sigma_\psi$=1.88'. For comparison, we give the error root-mean squares for non-mutual estimates: $\sigma_{o\Delta}$=2.8m; $\sigma_{o\psi}$=3.8×10$^{-3}$ 1/m (what corresponds to

$\sigma_{o\varphi}$=1.34').

**Conclusion.**

The analysis of SAF made allow judging principle potentialities of measuring of linear and angular (space-frequency) aperture position with respect to radiating background by the system of signal space processing. The results obtained may be a basis for evaluating of potentialities of correlation-extremum systems, which detect objects location simultaneously either by radiating background image or by background holographic image obtained in antenna aperture.

Two examples of SAF for continuous systems are considered in the paper. It is of interest to investigate ambiguity functions also for discrete systems of signal space processing working on the background of external radiations - multiple-beam systems, phased arrays, etc.

The consideration has been restricted by SAF analogous to frequency-time ambiguity function introduced in the works [1,2]. The analysis of generalized SAF may be made using the technique from [13,14]. For some additional information, one can also see [18].

**Table 1.**

| Problem Type | | Variant 1 | Variant 2 |
|---|---|---|---|
| Determination of Resolution | | $\delta_\Delta^{-2}=(2\pi)^2(l_{\psi f}^2+l_{\psi\sigma}^2+a^2 l_x^2)$ $\delta_\psi^{-2}=(2\pi)^2 l_x^2$ | $\delta_{\Delta_1}^{-2}=(2\pi)^2(l_{\psi\sigma}^2+a^2 l_x^2)$ $\delta_\psi^{-2}=(2\pi)^2 l_x^2$ |
| Determination of Measurements Accuracy | Non-mutual Measurements | $\sigma_{o\Delta}^{-2}=(2\pi)^2 q(l_{\psi f}^2+l_{\psi\sigma}^2+a^2 l_x^2)$ $\sigma_{o\psi}^{-2}=(2\pi)^2 q l_x^2$ | $\sigma_{o\Delta_1}^{-2}=(2\pi)^2 q(l_{\psi\sigma}^2+a^2 l_x^2)$ $\sigma_{o\psi}^{-2}=(2\pi)^2 q l_x^2$ |
| | Mutual Measurements | $\sigma_\Delta^{-2}=(2\pi)^2 q(l_{\psi f}^2+l_{\psi\sigma}^2)$ $\sigma_\psi^{-2}=\dfrac{(2\pi)^2 q(l_{\psi f}^2+l_{\psi\sigma}^2)l_x^2}{l_{\psi f}^2+l_{\psi\sigma}^2+a^2 l_x^2}$ $r_{\Delta\psi}=-\left[1+\dfrac{l_{\psi f}^2+l_{\psi\sigma}^2}{a^2 l_x^2}\right]^{-\frac{1}{2}}$ | $\sigma_{\Delta_1}^{-2}=(2\pi)^2 q l_{\psi\sigma}^2$ $\sigma_\psi^{-2}=\dfrac{(2\pi)^2 q l_{\psi\sigma}^2 l_x^2}{l_{\psi\sigma}^2+a^2 l_x^2}$ $r_{\Delta\psi}=-\left[1+\dfrac{l_{\psi\sigma}^2}{a^2 l_x^2}\right]^{-\frac{1}{2}}$ |

**FIGURE CAPTIONS.**

Fig. 1. Uncertainty in aperture linear displacement measuring: narrow (a) and wide (b) frequency-space spectrum of radiation.

Fig. 2. SAF and function (26) cross-sections for radiation far field zone.

Fig. 3. SAF and function (27) cross-sections for radiation near field region.

Fig. 4. Distribution of radiation energy $|\dot{\sigma}(\psi_x)|$ (a) and correlation function $K_b(\overline{\Delta})$ (b).

**REFERENCES**

[1] Woodward, P. M. (1955) Probability and Information Theory, with Applictions to Radar. New York: McGrow-Hill, 1955. London: Pergamon Press, 1955.

[2] Siebert, W. McC. (1956) A radar detection philosophy. The IRE Transactions on Information Theory, IT-2 (September 1956), No.3, p.204.

[3] Kelly, E. J. (1961) The radar measurements of range, velocity and acceleration. The IRE Transactions on Military Electronics, MIL-5 (April 1961), No.2, p.51.

[4] Urkowitz, H., Hauer, C. A., Koval, J. F. (1962) Generalized resolution in radar systems. Proceedings of the IRE, v.50 (October 1962), No.10, p.2093.

[5] Urkowitz, H. (1963) The angular ambiguity function of a discrete-continuous array. Proceedings of the IRE, v.51 (December 1963), No.12, p.1775.

[6] Urkowitz, H. (1964) The accuracy of maximum likelyhood angle estimates in radar and sonar. The IRE Transactions on Military Electronics, MIL-8 (January 1964), No.1, p.39.

[7] Vakman, D. E. (1965) Compecated Signals and Ambiguity Principle in Radar. Moscow: Soviet Radio, 1965 (in Russian).

[8] Shirman, Ya. D., Manzhos, V. H. (1981) Theory and Technology of Radar Information Processing on Interferences Background. Moscow: Radio and Communications, 1981 (in Russian).

[9] Reutov, A. P., Mikhaylov, B. A., Kondratenkov, G. S., Boykov, B. V. (1970) Lateral Scanning Radars. Moscow: Soviet Radio, 1970 (in Russian).

[10] Space-Time Signal Processing (1984) Ed. by Kremer, I. Ya. Moscow: Radio and Communications, 1981 (in Russian).

[11] Karavayev, V.V., Sazonov, V.V. (1982) On signal detection when synthesizing the aperture in spaced systems. Radioteknika i Elektronika, v.27 (June 1982), No.6, p.1132 (in Russian).

[Engl. Transl.: Soviet Journal of Communication Technology and Electronics, v.27 (June 1982), No.6].

[12] Young, G. O., Howard, J. E. (1970) Application of space-time decision and estimation theory to antenna system design. Proceedings of the IRE, v.58 (May 1970), No.5, p.771.

[13] Yuryev, A. N. (1972) Accuracy of mutual estimation of the carrier frequency and the direction of radiosignal coming. Radioteknika i Elektronika, v.17 (February 1972), No.2, p.301 (in Russian).

[Engl. Transl.: Soviet Journal of Communication Technology and Electronics, v.17 (February 1972), No.2].

[14] Yuryev, A. N., Muravyev, P. A. (1974) The accuracy of signal parameters estimation in the case of focused antenna using. Radioteknika i Elektronika, v.19 (March 1974), No.4, p.697 (in Russian).

[Engl. Transl.: Soviet Journal of Communication Technology and Electronics, v.19 (March 1974), No.4].

[15] Krasovski, A. A., Beloglazov, I. N., Chigin, G. P. (1979) The Theory of Correlation-Extremum Systems. Moscow: Nauka, 1979 (in Russian).

[16] Baklitski, V. K., Yuryev, A. N. (1982) Correlation-Extremum Methods of Navigation. Moscow: Radio and Communications, 1982 (in Russian).

[17] Falkovich, S. E. (1974) Signal Parameters Estimation. Moscow: Soviet Radio, 1974 (in Russian).

[18] Yuryev, A. N. (1989) Simultaneous estimation of signal space parameters from extended targets. Radioelektronika, v.32 (May 1989), No.5, p.12 (in Russian).

[Engl. Transl.: Radioelectronics and Communications Systems, v.32 (May 1989), No.5].

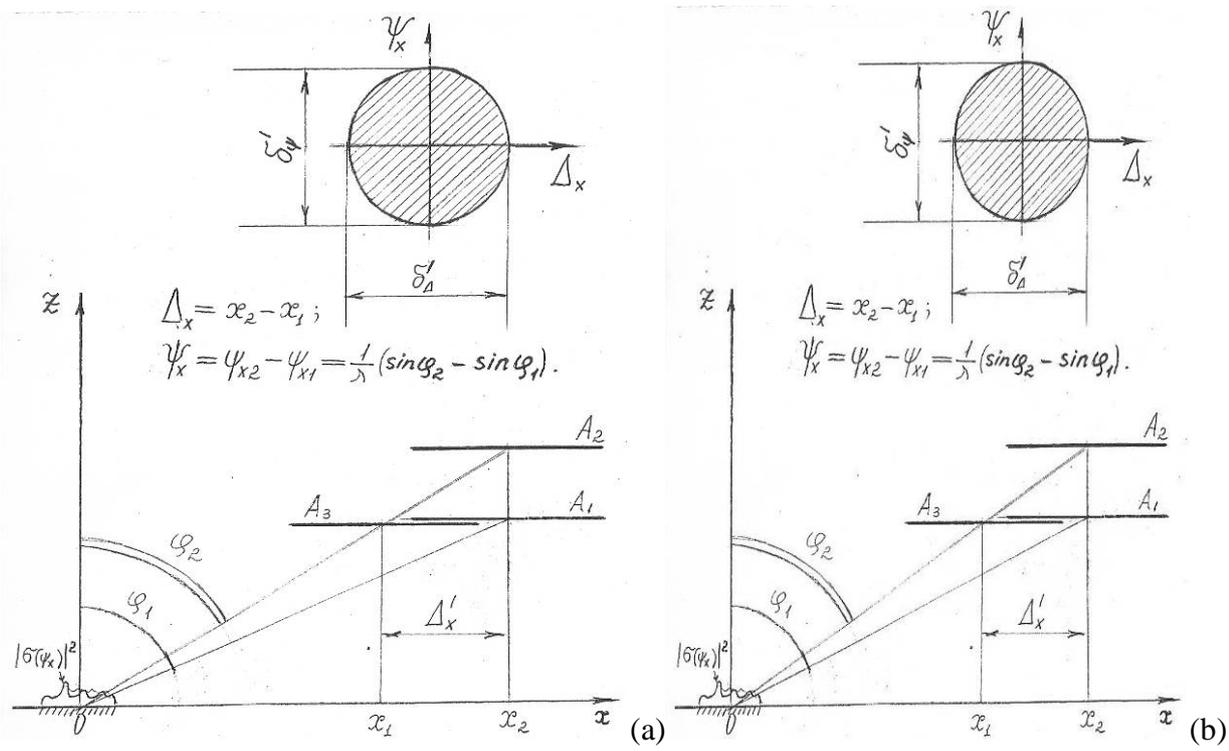

Fig. 1.

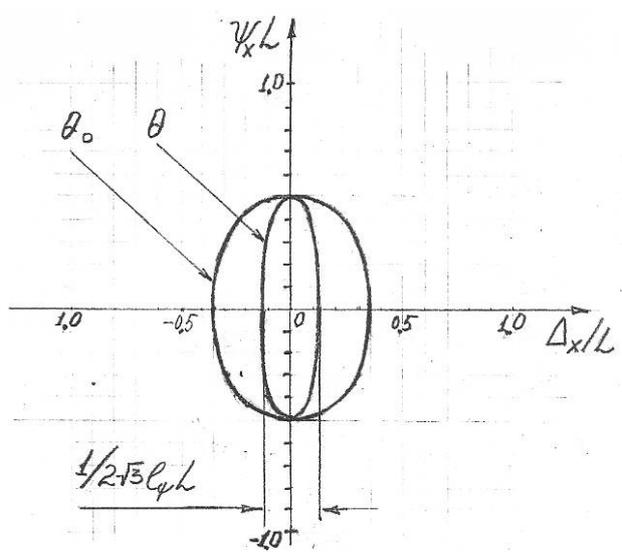

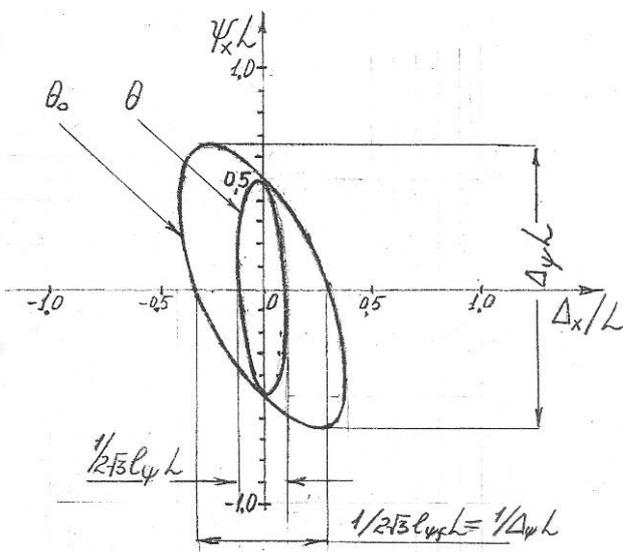

Fig. 2.                                      Fig. 3.

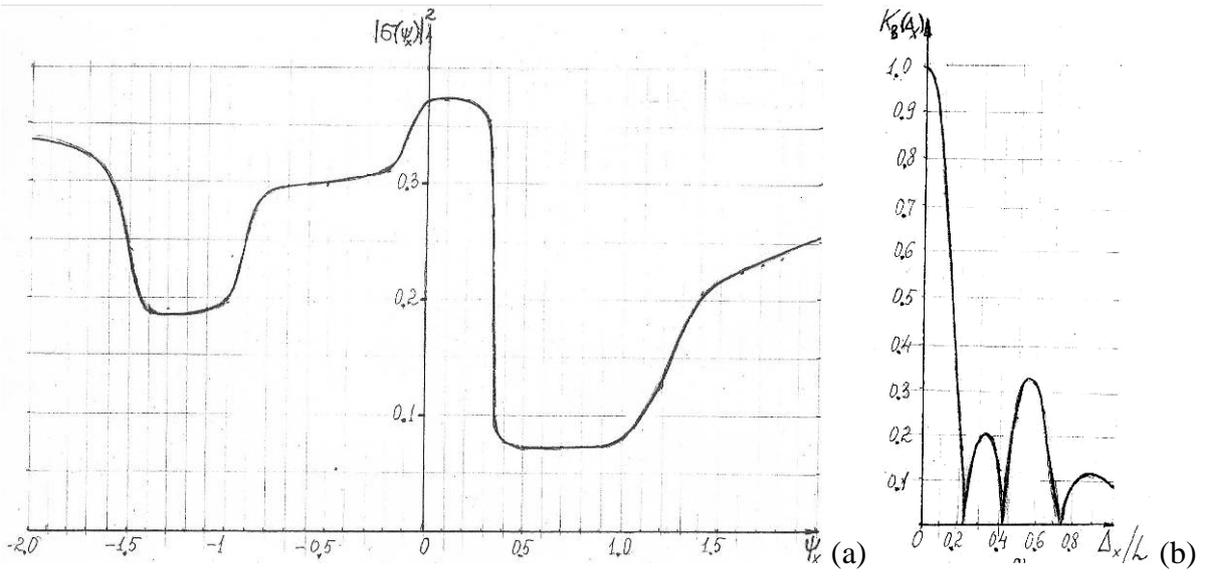

Fig. 4.